\documentclass[%
 reprint,
 superscriptaddress,
 showkeys, 
 10pt,
 twocolumn,
 ]{revtex4-2}
\renewcommand{\section}[1]{\textit{#1.--}} 

\usepackage[utf8]{inputenc}			
\usepackage[british,english]{babel}	        
\usepackage{amsmath}				
\usepackage{amssymb}				
\usepackage{mathtools}				
\usepackage{bm}						
\usepackage{epsfig}
\usepackage{graphicx}				
\usepackage{xcolor} 				
\usepackage{comment} 				
\usepackage{array}					
\usepackage{tabularx}				
\usepackage[english=british]{csquotes} 
\usepackage{orcidlink}              
\usepackage{ragged2e}               

\usepackage{hyperref}		
\hypersetup{hidelinks} 		
\hypersetup{
    colorlinks = true,
    urlcolor   = blue,
    citecolor  = blue,
}

\usepackage{dcolumn}

\renewcommand{\Pr}		{\mathrm{Pr}} 

\renewcommand{\Re}		{\mathrm{Re}} 
\newcommand{\Ri}		{\mathrm{Ri}}

\newcommand{\dO}		{d_{0}}

\newcommand{\UO}		{U_{0}}
\newcommand{\BO}		{B_{0}}

\newcommand{\ReO}		{\mathrm{Re}_{0}}
\newcommand{\RiO}		{\mathrm{Ri}_{0}}

\newcommand{\Rig}		{\mathrm{Ri}_{\textrm{g}}}

\newcommand{\tstar}		{t_{\star}}
\newcommand{\zstar}		{z_{\star}}

\newcommand{\tcrit}		{t_{\textrm{del}}}

\makeatletter
\newcommand\thefontsize{The current font size is: \f@size pt}
\makeatother

\begin{document}
\title{Stability of Diffusive Shear Layers}

\author{Stefan S. Nixon}
  \affiliation{Department of Applied Mathematics and Theoretical Physics, Wilberforce Road, Cambridge, CB3 0WA, United Kingdom.}
  \affiliation{Department of Aeronautics, Imperial College London, South Kensington Campus, London, UK}
\author{Philipp P. Vieweg \orcidlink{0000-0001-7628-9902}}
  \email{Contact author: ppv24@cam.ac.uk}
  \affiliation{Department of Applied Mathematics and Theoretical Physics, Wilberforce Road, Cambridge, CB3 0WA, United Kingdom.}

\date{\today}

\begin{abstract}
As one of the cornerstones of fluid mechanics, stability analyses provide essential physical insights into the growth of perturbations and eventual transition to turbulence.
However, classical \enquote{frozen-time} stability analyses implicitly assume a time-independence of their base flow and thus fail for \enquote{rapidly} diffusing shear layers. 
Here, we propose a self-similar ansatz to naturally incorporate the \enquote{diffusive} base-state expansion into the stability operator. 
Our approach reveals two competing physical mechanisms: an \enquote{expansion wind} delays the Kelvin-Helmholtz instability whereas a diminishing effective viscosity sustains this instability far beyond classical predictions. 
Direct numerical simulations confirm that our framework accurately captures the instability's extended lifespan, growth rate, and spectral topology, eventually revising the timeline of shear-induced mixing fundamentally.
\end{abstract}

\maketitle
\section{Introduction}
The stability of shear flows represents a fundamental problem in fluid dynamics and serves as a primary mechanism for mixing in a wide range of flows such as planetary boundary layers \cite{newsom2003,Thorpe2007} and industrial flows \cite{liu2015kelvin,bovcek2024kelvin}. Classical inviscid theory dictates that stratified shear flows may become unstable when the gradient Richardson number $\Rig < 1/4$ \citep{Miles1961,Howard1961}. However, in physical reality, finite-thickness shear layers continuously diffuse. The stability boundary is no longer singular but depends on a complex interplay between the Prandtl ($\Pr$), Reynolds ($\Re$), and Richardson ($\Ri$) numbers. Despite its ubiquity, the role of base-state evolution -- i.e., the \enquote{spreading} of the shear layer -- is often neglected in quasi-steady stability analyses. In this Letter, we show that the time-dependent expansion of the base state is not a mere perturbation but a primary determinant of the Kelvin-Helmholtz instability (KHI). We demonstrate that this expansion governs the selection of unstable modes and the subsequent transition to turbulence, providing a new framework for predicting the onset of instability in evolving shear layers.

\section{System Formulation}
\label{sec:Setup}
We consider a miscible, incompressible fluid governed by the Oberbeck-Boussinesq Navier-Stokes equations, non-dimensionalised by the initial streamwise velocity $\UO$, buoyancy $\BO$, and shear layer half-depth $\dO$. Hence, the dimensionless evolution of the velocity $\bm{u}$, buoyancy $b$, and modified pressure $p$ is governed by
\begin{align}
\label{eq:NSE}
\frac{\partial \bm{u}}{\partial t} + \left( \bm{u} \cdot \nabla \right) \bm{u} &= - \nabla p + \frac{1}{\ReO} \nabla^{2} \bm{u} + \RiO b \bm{e}_{z}, \\
\label{eq:BE}
\frac{\partial b}{\partial t} + \left( \bm{u} \cdot \nabla \right) b &= \frac{1}{\ReO \Pr} \nabla^{2} b ,
\end{align}
subject to the continuity constraint $\nabla \cdot \bm{u} = 0$. The system is naturally characterised by the Prandtl number $\Pr := \nu/\kappa$, initial bulk Reynolds number $\ReO := \UO \dO/\nu$, and initial bulk Richardson number $\RiO := \BO \dO/\UO^{2}$. We restrict our attention to the statically stable regime ($\RiO > 0$), with specific focus on $\Pr = 1$, $\ReO = 50$, and $\RiO = 0.0125$. Direct numerical simulations of a similar system have been conducted by \citet{Vieweg2026}.

\section{Diffusive stability analysis}
\label{sec:Diffusive_stability_analysis}
Classically, the linear stability of a shear layer is analysed by decomposing the flow variables $\bm{\Phi}$ into a static base state and a tiny perturbation via the standard normal-mode ansatz $\bm{\Phi}(x, z, t) = \bar{\bm{\Phi}}(z) + \hat{\bm{\Phi}}(z) e^{i(kx - \sigma t)}$. 
To account for diffusive spreading, i.e. $\bar{\bm{\Phi}} = \bar{\bm{\Phi}}(z, t)$, the quasi-steady or \enquote{frozen-time} approximation is typically employed \citep{drazin2004hydrodynamic,Smyth2011}. This implicitly assumes the diffusive timescale far exceeds the instability's e-folding time, treating the base state as steady at distinct snapshots. The resulting ansatz
\begin{equation}
\label{eq:standard_ansatz_mod}
    \bm{\Phi}(x, z, t; t_{0}) = \bar{\bm{\Phi}}(z, t) + \hat{\bm{\Phi}}(z) e^{i(kx - \int_{t_0}^t \sigma(t') dt')}
\end{equation}
is beholden to this slow diffusive time.
However, for particularly fine interfaces (where $\dO \to 0$), this separation of timescales breaks down: The diffusive timescale becomes comparable to or smaller than the instability's growth time, rendering the frozen-time assumption invalid. 

Relying on $\dO$ for non-dimensionalisation becomes obstructive when the initial interface approaches a step function because the associated $\ReO \to 0$ limit causes a mathematical singularity in the equations of motion. Hence, we introduce an alternative diffusive scaling, indicated by the subscript $_{\star}$, based on fixed molecular properties (see End Matter). 

To accommodate the strong coupling between the (rapidly) evolving background flow and the perturbations, we employ a self-similar framework analogous to methods developed for viscous fingering \citep{Pramanik2013,Pramanik2015}. By applying the mapping $\left( \zstar, \tstar \right) \mapsto \left( \xi = \zstar / \sqrt{\tstar}, \tstar \right)$, the base states become steady in the new coordinate $\xi$, allowing the base flow to evolve as tractable, self-similar diffusing profiles given by
\begin{equation}
\label{eq:diffusive_profiles}
    U(\xi, \tstar) = \operatorname{erf}\left( \frac{\xi}{2 \sqrt{\Pr}} \right), \quad
    B(\xi,\tstar) = \operatorname{erf}\left( \frac{\xi}{2} \right).
\end{equation}

Instead of assuming a separation of timescales, this coordinate transformation naturally absorbs the diffusive growth directly into the governing operators. We then seek normal mode solutions taking the fundamentally distinct form
\begin{equation}
\label{eq:ansatz_dif}
\bm{\Phi} \left( \xi, \tstar \right) = \bar{\bm{\Phi}} \left( \xi \right) + \hat{\bm{\Phi}} \left( \xi \right) \tau \left( \tstar \right) e^{ikx} .
\end{equation}
Unlike the standard frozen-time approach, this self-similar \enquote{diffusive} ansatz introduces a one-way coupling where the temporal variation of the perturbation, $\tau(\tstar)$, evolves in tandem with the expanding background flow.

Solving the resulting self-similar Taylor-Goldstein equations yields the instantaneous growth rate $\sigma(\tstar) = \tau^{-1}(d\tau/d\tstar)$. Crucially, this formulation allows us to define the cumulative growth
\begin{equation}
\label{eq:cumulative_growth}
    \mathcal{G}(t; t_{0}) := \frac{\tau(t)}{\tau(t_0)} = \exp\!\left(\int_{t_0}^{t} \sigma(t')\, dt' \right)
\end{equation}
over a finite interval $(t_0, t)$ as the ratio of temporal amplitudes which integrates the temporal history of the instability and captures the dynamic lag introduced by finite-rate diffusion.

\begin{figure}[t]
\centering
\includegraphics[scale = 1.0]{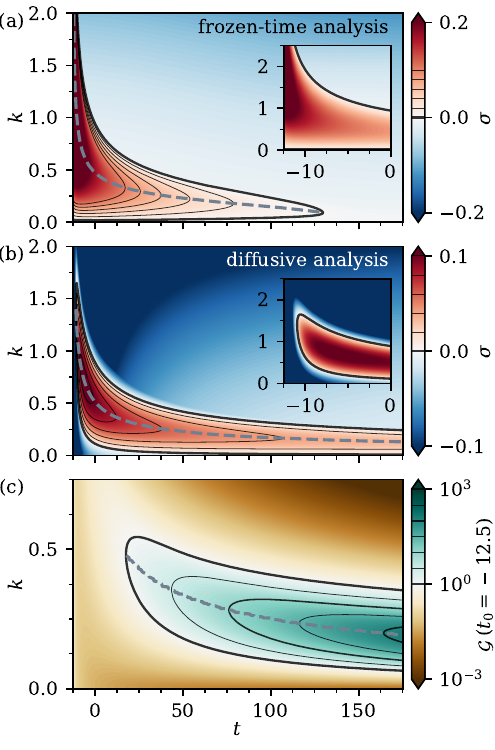}
\caption{\justifying{
Theoretical modal stability and growth. 
In contrast to (a) the frozen-time analysis, 
the (b) instantaneous growth rate $\sigma$ from the diffusive approach highlights a delayed onset at early times and an extended persistence of instabilities at late times. 
(c) The associated cumulative growth $\mathcal{G} \left( k, t, t_{0} = - \ReO / 4 \right)$ demonstrates the severe initial attenuation and the subsequent prolonged recovery phase required to achieve net perturbation amplification. 
Dashed lines track maxima.
}}
\label{fig:stability_and_growth}
\end{figure}

\section{Physical Mechanisms Governing Diffusive Stability}
\label{sec:Physical_Mechanisms_Governing_Diffusive_Stability}
The translation to the self-similar coordinate $\xi$ introduces non-autonomous terms scaling with $t^{-1}$ into the linear evolution operator (matrix $\bm{A}$, see End Matter). These terms reveal two competing physical mechanisms that fundamentally alter the stability window compared to the classical frozen-time approach.

First, the pseudo-advective term $\left( \xi / 2t \right) D_\xi$ (where $D_\xi$ is the derivative with resect to $\xi$) acts as an \enquote{expansion wind} driven by the spreading coordinate system. At early times ($t \to 0$), this rapid base-state stretching dilutes perturbation energy and suppresses the resonant interaction required for shear instability. Consequently, the linear onset of instability is significantly delayed [Fig.~\ref{fig:stability_and_growth}(a,b)]. The practical impact of this early-time suppression is quantified by the cumulative growth $\mathcal{G}$ [Fig.~\ref{fig:stability_and_growth}(c)]: The initial stable phase causes substantial perturbation decay, requiring a prolonged recovery period; perturbations attenuated during the first $3$ time units require a further $\approx 30$ time units of growth merely to restore their initial energy level. 

Second, at late times, viscous dissipation scales as $\Pr \nabla^2 \sim (\Pr/t) D_\xi^2$ in the transformed frame, so this dissipative term decays as the shear layer thickens ($t \to \infty$). Physically, the effective Reynolds number $\Re_{\textrm{eff}} = \UO d_{\textrm{dim}}(t) / \nu \propto \sqrt{t}$ grows, driving the evolving flow asymptotically toward an inviscid regime. Ultimately, this diminishing effective viscosity dynamically sustains the instability even as the bulk velocity gradients weaken [Fig.~\ref{fig:stability_and_growth}(a,b), $t>100$], whereas a classical frozen-time analysis prematurely predicts restabilisation.

Ultimately, the stability of the diffusing interface is governed by this dynamic competition: the expansion wind delays the initial macroscopic onset, while the diminishing effective viscosity extends the late-time unstable window.

\section{Numerical validation}
\label{sec:Numerical_validation}
To rigorously validate our diffusive approach, we confront it with three-dimensional direct numerical simulations (DNS). We solve the governing equations using spectral element methods \citep{Fischer2022, Vieweg2026} in a Cartesian domain of size $L_x = L_y = 512$ and $L_z = 48$. 

As true step-functions are computationally prohibitive, we follow the established framework \citep{Smyth2000, Smyth2011, Smith2021, Vieweg2026} of initializing the simulation at $t = 0$ with $d = 1$, corresponding to base profiles $u_{x, 0} = \operatorname{erf}(z)$ and $b_{0} = \operatorname{erf}(\sqrt{\Pr} \thickspace z)$. To seed instabilities without artificially prescribing selected structures, we introduce random buoyancy fluctuations bounded by $\pm 10^{-3}$. In the context of our diffusive theory, this specification of $t=0$ corresponds to a theoretical origin of $t = -\ReO/4 = -12.5$ for a step-function interface. Thus, the initial \enquote{expansion wind} phase predicted in Fig.~\ref{fig:stability_and_growth} occurs prior to simulation start.

\begin{figure}[t]
\centering
\includegraphics[scale = 1.0]{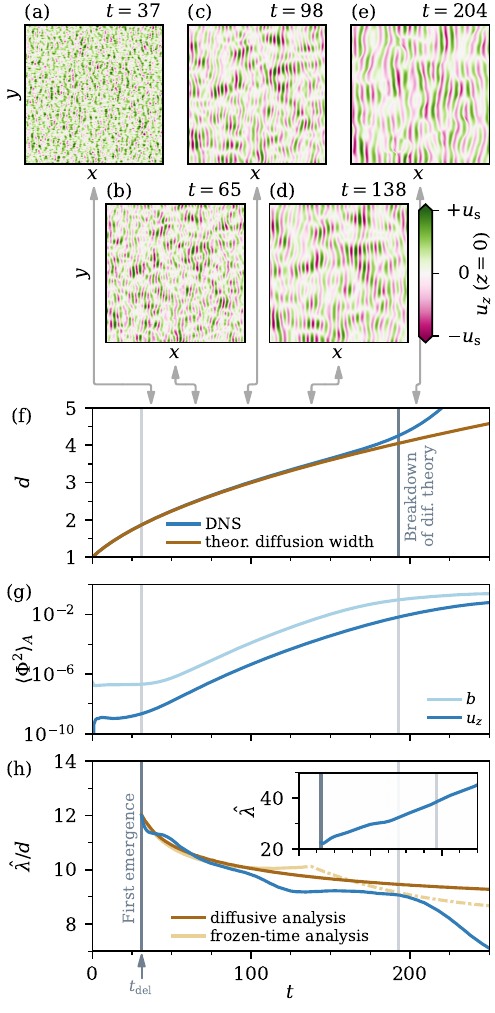}
\caption{\justifying{Fully non-linear DNS. 
(a--e) Coherent structures 
(colour saturated at $u_{\textrm{s}} = 3\langle u_{z}^{2} \rangle_{A}^{1/2}$) 
manifest in tandem with a 
(f) spreading shear layer and
(g) growth of variance at midplane, 
revealing the 
(g) initial stochastic induction phase and 
(f) eventual onset of strong nonlinearity.
The competition between spreading and modal growth governs 
(h) the dominant normalised instability wavelength. 
}}
\label{fig:direct_numerical_solution}
\end{figure}

As the flow evolves, coherent structures emerge and amplify [Fig.~\ref{fig:direct_numerical_solution}(a--e)]. To quantify the base-state broadening, we track the momentum (half-)thickness $d(t)~:=~\sqrt{\pi/8} \thickspace \int \left( 1 - \langle u_x \rangle_A^2 \right) dz$ \cite{Smyth2000} with $\langle \cdot \rangle_A$ denoting a horizontal average. Figure~\ref{fig:direct_numerical_solution}(f) demonstrates excellent agreement between the computed $d(t)$ and the theoretical self-similar prediction $d_{\textrm{th}}(t) = \sqrt{1 + 4t/\ReO}$ for an extended period, diverging only at $t \sim \mathcal{O}(10^2)$ upon entering the strongly nonlinear regime \cite{Arratia2013}. 

Prior to exponential growth, the temporal evolution of the variance in $b$ and $u_z$ [Fig.~\ref{fig:direct_numerical_solution}(g)] exhibits a distinct quasi-steady induction period which reflects the stochastic initialisation \citep{frame2024beyond}. While the variance is initially dominated by a superposition of decaying or neutral modes across the eigenspectrum, this period (of $\tcrit \approx 30$) represents the statistical delay required for the primary unstable eigenmode to amplify sufficiently to overcome the noise floor of the stable spectrum and dominate the macroscopic flow.

Crucially, the dominant instability length scale [Fig.~\ref{fig:direct_numerical_solution}(h)] captures this dynamic competition. Because diffusive interface spreading outpaces the physical wavelength shift, the normalised peak wavelength decreases over time. When compared against theoretical predictions, both the frozen-time and diffusive models initially capture the trend exceptionally well. 
However, a clear bifurcation occurs at late times. 
The classical frozen-time analysis fails to capture the continued evolution of the structures as it prematurely predicts stabilisation (dash-dotted). 
In stark contrast, our diffusive analysis successfully tracks the sustained growth of the instability deep into the late-time regime, validating the \enquote{diminishing effective viscosity} mechanism right up until nonlinear breakdown.

\begin{figure}[t]
\centering
\includegraphics[scale = 1.0]{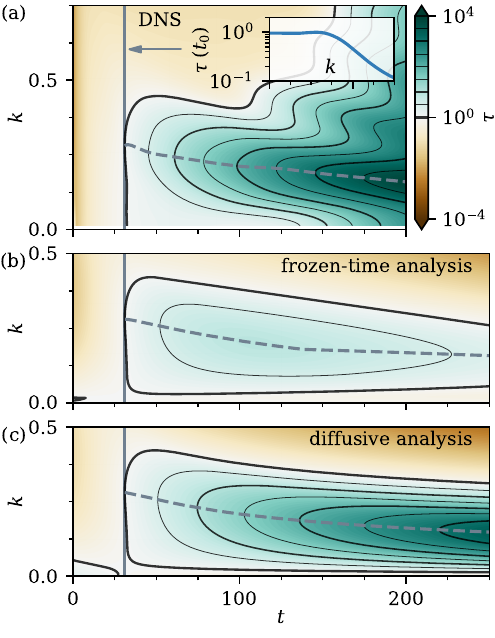}
\caption{\justifying{
Evolution of amplitude spectra. 
(a) The DNS amplitude spectrum exposes the spectral peak's gradual shift (dashed line) and successive three-dimensional forward cascade.
Its instantaneous $\tau \left( t_{0} = \tcrit \right)$ (inset) is used to initialise the theoretical models.
(c) The diffusive analysis accurately predicts the amplitude magnitudes leading to nonlinear breakdown, whereas (b) the frozen-time analysis significantly underestimates resulting growth.
}}
\label{fig:growth_of_amplitudes}
\end{figure}

\section{Spectral Evolution and Amplitude Growth}
\label{sec:Spectral_Evolution_and_Amplitude_Growth}
We track the evolution of the normalised DNS amplitude spectrum, $\tau(k, t) = A(k, t) / \max_{k} A(k,\tcrit)$, in Fig.~\ref{fig:growth_of_amplitudes}(a). The dashed line follows the spectral peak, representing the mode of greatest accumulated growth. As the shear layer coarsens, this peak amplifies and shifts toward lower wavenumbers. 
At late times ($t \gtrsim 100$), energy simultaneously accumulates in high-wavenumber modes. This signals the onset of a three-dimensional forward cascade \cite{Verma2018} of kinetic energy -- a nonlinear turbulent breakdown that falls outside the scope of linear theory \cite{mashayek2012a,mashayek2012b,salehipour2015turbulent}.

Guided by the physical limits established in Fig.~\ref{fig:direct_numerical_solution}, we restrict our theoretical comparison to the linear regime $t \ge \tcrit$. We initialise both theoretical models using $\tau(k, t_{0} = \tcrit)$ from the DNS [Fig.~\ref{fig:growth_of_amplitudes}(a), inset] and compute the subsequent spectral evolution via $\mathcal{G}(k, t; t_{0} = \tcrit)$.

While both the frozen-time and diffusive approach capture the temporal shift of the peak wavenumber with reasonable accuracy [Fig.~\ref{fig:growth_of_amplitudes}(b, c)], the former severely underestimates the cumulative amplitude growth. 
This underprediction of the amplitude $A \propto \tau$ spuriously suggests that the linear regime, whose lifespan is governed by the wave steepness criterion $A k \ll 1$ \cite{drazin2004hydrodynamic}, persists much longer than physically observed. 
In contrast, the fully diffusive theory accurately captures the magnitude of the growth, providing a highly reliable, mathematically tractable prediction for the threshold of nonlinear breakdown.

\begin{figure}[t]
\centering
\includegraphics[scale = 1.0]{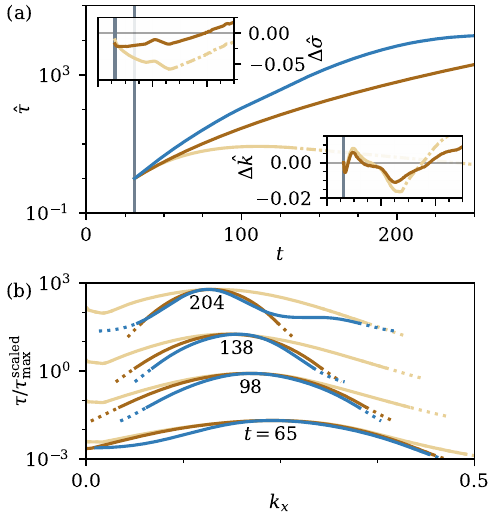}
\caption{\justifying{
Growth and topology of the emergent dominant mode. 
In contrast to the frozen-time analysis (which spuriously predicts premature attenuation), the diffusive analysis accurately predicts the (a) sustained growth of the peak amplitude and (b) associated spectral topology (incl. the high-wavenumber asymmetry) deep into the late-time regime
due to (a, insets) significantly reduced errors of the peak's growth rate and wavenumber.
In (b), spectra are vertically offset by $\tau_{\textrm{max}}^{\textrm{scaled}}=\max_{k} (\tau) \thickspace 10^{18} / t^{9}$ for visual clarity.
See Fig.~\ref{fig:direct_numerical_solution}(f--h) for colour encoding.
}}
\label{fig:evolution_peak_amplitude}
\end{figure}

Isolating the temporal evolution of the dominant spectral peak further exposes the fundamental breakdown of the classical frozen-time approach. While the diffusive approach robustly recovers the monotonic amplification of the peak amplitude $\hat{\tau}$ observed in the DNS [Fig.~\ref{fig:evolution_peak_amplitude}(a)], the frozen-time approach suffers a qualitative failure. Rather than correctly predicting sustained growth, the frozen-time analysis prematurely predicts the stabilisation of the instability, resulting in a spurious cusp in the amplitude evolution. 
This structural discrepancy is rigidly quantified by the instantaneous growth rate error, $\Delta \hat{\sigma} = \hat{\sigma}_{\Phi} - \hat{\sigma}_{\textrm{DNS}}$ [Fig.~\ref{fig:evolution_peak_amplitude}(a), upper inset]. Here, the diffusive ansatz achieves a more than three-fold reduction in mean error ($\Delta \hat{\sigma} = -0.014 \pm 0.006$) compared to the prematurely stabilised frozen-time prediction ($\Delta \hat{\sigma} = -0.044 \pm 0.010$), demonstrating its systematically superior accuracy for non-stationary base flows.

Evaluating the corresponding peak wavenumber $\hat{k}$ reveals a broader consensus between the models, though structural differences remain. Both approaches capture the macroscopic shift toward lower wavenumbers as the shear layer coarsens, a trend previously established in Fig.~\ref{fig:direct_numerical_solution}(h). The deviations $\Delta \hat{k}$ [Fig.~\ref{fig:evolution_peak_amplitude}(a), lower inset] remain quantitatively small for both methods, however this superficial similarity masks a fundamental mechanistic divergence. The physical consistency of the diffusive analysis stems from its ability to correctly predict the continued growth of the instability in line with the DNS. In contrast, the frozen-time approach initially overpredicts $\hat{k}$ before prematurely predicting stabilisation; as the mode artificially dies in this model, $\hat{k}$ coincidentally drops back toward the DNS values. Despite this deceptive numerical proximity, the late-time fidelity of the diffusive approach is physically paramount: as the instability length scale diverges ($\hat{k} \to 0$), even marginal inaccuracies in $\hat{k}$ translate into severe, compounding deviations in the predicted physical wavelength, $\hat{\lambda} \propto \hat{k}^{-1}$.

Even beyond integral metrics, the diffusive theory faithfully preserves the underlying spectral topology [Fig.~\ref{fig:evolution_peak_amplitude}(b)] across the developmental stages corresponding to Fig.~\ref{fig:direct_numerical_solution}(b--e). It accurately captures the spectral shape, particularly the structural asymmetry in the high-wavenumber tail ($k \gtrsim \hat{k}$). Remarkably, this morphological agreement with the DNS persists deep into the late-time regime, confirming that the self-similar diffusive approach effectively captures the essential physics of modal selection in evolving shear layers.

\section{Final remarks}
\label{sec:final_remarks}
In this Letter, we have demonstrated that the classical frozen-time approximation fundamentally misrepresents the linear stability of rapidly diffusing stratified shear layers. By employing a self-similar coordinate transformation, our diffusive stability framework naturally incorporates the time-dependent expansion of the base states. 
In essence, this approach respects that the interface \textit{is diffusing} rather than that it just \textit{is diffuse}.
Our formulation isolates the physical mechanisms governing the transition to turbulence: an initial \enquote{expansion wind} suppresses perturbation growth and delays the macroscopic onset of instability, whereas a diminishing effective viscosity dynamically sustains this instability deep into the late-time evolution. Direct numerical simulations unequivocally validate this framework, confirming that the diffusive ansatz accurately captures the cumulative growth, scale evolution, and spectral morphology of the emergent dominant modes.

Unsurprisingly, a singular analytical stability bound -- analogous to the classical inviscid Miles-Howard criterion \cite{Miles1961, Howard1961}, $\Rig < 1/4$ -- cannot be derived for the fully diffusive system. Evaluating the self-similar base states at the interface core shows that the local gradient Richardson number, $\Rig(\zstar = 0, \tstar) = \Ri_\kappa \Pr \thickspace \sqrt{\pi \tstar}$, is strictly time-dependent. While $\Rig$ is vanishingly small at early times (which classically favours instability), perturbation growth is instead overpowered by the expansion wind. Conversely, at late times, continuous diffusive thickening eventually drives $\Rig$ beyond the critical threshold. Consequently, the transition to turbulence must be evaluated as a transient, dynamic trajectory governed by the history of the evolving base states, rather than crossing a static scalar boundary.

Ultimately, this diffusive stability framework provides a physically rigorous timeline for the lifespan of the linear regime, reinforcing that base-state evolution is a primary determinant of instability rather than a mere perturbation. In broad physical contexts -- ranging from planetary boundary layers \cite{newsom2003,Thorpe2007} to industrial shear flows \cite{liu2015kelvin,bovcek2024kelvin} -- relying on classical, quasi-steady stability criteria can lead to drastic miscalculations in the timing and efficiency of mixing events. By accurately capturing the precise delay, duration, and spectral topology of shear-induced mixing under the evolving interplay of $\Pr$, $\Re$, and $\Ri$, our time-dependent approach paves the way for more faithful sub-grid parameterisations in large-scale climate and engineering fluid models.

\section{Data Availability}
The data that support the findings of this Letter -- including MATLAB and Python codes to solve the diffusive stability analysis -- will be made openly available after acceptance of the latter.

\section{Acknowledgments}
The authors thank Stuart B. Dalziel and Colm-cille P. Caulfield (University of Cambridge) as well as Marc Avila (University of Bremen) for a fruitful early discussion.
P.P.V. gratefully acknowledges the support of Pembroke College, Cambridge, through a postdoctoral research associateship and is funded by the Deutsche Forschungsgemeinschaft (DFG, German Research Foundation) within Walter Benjamin Programme 532721742. 
The authors gratefully acknowledge 
(i) the Gauss Centre for Supercomuting e.V. (\href{www.gauss-centre.eu}{\texttt{www.gauss-centre.eu}}) for funding this work by providing computing resources through the John von Neumann Institute for Computing (NIC) on the GCS supercomputer JUWELS at Jülich Supercomputing Center (JSC) within project nonbou and 
(ii) the Cirrus UK National Tier-2 HPC Service at EPCC (\href{www.cirrus.ac.uk}{\texttt{www.cirrus.ac.uk}}) -- funded by The University of Edinburgh, the Edinburgh and South East Scotland City Region Deal, and UKRI via EPSRC -- for supporting this work through project e970.

\section{Author Contributions}
Both authors designed the study. 
S.S.N. derived the diffusive stability analysis and solved it numerically. 
P.P.V. conducted the numerical simulations, merged, and processed the generated data. 
Both authors contributed equally in discussing the data and writing the paper.


\bibliography{references}

@PREAMBLE{
 "\providecommand{\noopsort}[1]{}" 
 # "\providecommand{\singleletter}[1]{#1}%" 
}

@article{Smith2021, 
title={Turbulence in forced stratified shear flows}, volume={910}, 
journal={J. Fluid Mech.}, 
author={Smith, Katherine M. and Caulfield, C.P. and Taylor, J.R.}, 
year={2021}, 
pages={A42}}

@article{Pramanik2013,
issn = {1070-6631},
journal = {Phys. Fluids},
volume = {25},
publisher = {Amer. Inst. Physics},
number = {7},
year = {2013},
title = {Linear stability analysis of {Korteweg} stresses effect on miscible viscous fingering in porous media},
copyright = {Copyright 2019 Elsevier B.V., All rights reserved.},
address = {MELVILLE},
author = {Pramanik, Satyajit and Mishra, Manoranjan},
}

@article{Pramanik2015,
issn = {0009-2509},
journal = {Chem. Eng. Sci.},
pages = {523--532},
volume = {122},
publisher = {Elsevier Ltd},
year = {2015},
title = {Nonlinear simulations of miscible viscous fingering with gradient stresses in porous media},
copyright = {2014 Elsevier Ltd},
address = {OXFORD},
author = {Pramanik, Satyajit and Mishra, Manoranjan},
}

@article{Miles1961,
  title={On the stability of heterogeneous shear flows},
  author={Miles, John W},
  journal={J. Fluid Mech.},
  volume={10},
  number={4},
  pages={496--508},
  year={1961},
  publisher={Cambridge University Press}
}

@article{Smyth2011,
  title={Narrowband oscillations in the upper equatorial ocean. {Part} II: {Properties} of shear instabilities},
  author={Smyth, WD and Moum, JN and Nash, JD},
  journal={J. Phys. Oceanogr.},
  volume={41},
  number={3},
  pages={412--428},
  year={2011}
}

@article{frame2024beyond,
  title={Beyond optimal disturbances: a statistical framework for transient growth},
  author={Frame, Peter and Towne, Aaron},
  journal={J. Fluid Mech.},
  volume={983},
  pages={A2},
  year={2024},
  publisher={Cambridge University Press}
}

@article{Vieweg2026,
  title={Anisotropy of emergent large-scale dynamics in forced stratified shear flows},
  author={Vieweg, P.P. and Caulfield, C.P.},
  journal={J. Fluid Mech.},
  volume={1029},
  pages={A7},
  year={2026},
  publisher={Cambridge University Press}
}

@article{mashayek2012a,
  title={The ‘zoo’ of secondary instabilities precursory to stratified shear flow transition. {Part} 1: {Shear} aligned convection, pairing, and braid instabilities},
  author={Mashayek, A and Peltier, WR},
  journal={J. Fluid Mech.},
  volume={708},
  pages={5--44},
  year={2012},
  publisher={Cambridge University Press}
}

@article{mashayek2012b,
  title={The ‘zoo’of secondary instabilities precursory to stratified shear flow transition. {Part} 2: {The} influence of stratification},
  author={Mashayek, A and Peltier, WR},
  journal={J. Fluid Mech.},
  volume={708},
  pages={45--70},
  year={2012},
  publisher={Cambridge University Press}
}

@article{salehipour2015turbulent,
  title={Turbulent diapycnal mixing in stratified shear flows: {The} influence of {Prandtl} number on mixing efficiency and transition at high {Reynolds} number},
  author={Salehipour, H. and Peltier, W.R. and Mashayek, A.},
  journal={J. Fluid Mech.},
  volume={773},
  pages={178--223},
  year={2015},
  publisher={Cambridge University Press}
}

@article{Fischer2022,
  title={{NekRS}, a {GPU}-accelerated {Spectral} {Element} {Navier–Stokes} {Solver}},
author={Fischer, P. and Kerkemeier, S. and Min, M. and Lan, Y.-H. and Phillips, M. and Rathnayake, T. and Merzari, E. and Tomboulides, A. and Karakus, A. and Chalmers, N. and Warburton, T.},
  journal={Parallel Comput.},
  volume={114},
  pages={102982},
  year={2022},
  publisher={American Meteorological Society}
}

@Article{Smyth2000,
  author = 	 {Smyth, W.D. and J.N. Moum},
  title = 	 {Length scales of turbulence in stably stratified mixing layers},
  journal = 	 {Phys. Fluids},
  year = 	 {2000},
  volume = 	 {12},
  pages = 	 {1327--1342},
}

@book{Verma2018,
  title = {Physics of {{Buoyant Flows}}: {{From Instabilities}} to {{Turbulence}}},
  shorttitle = {Physics of Buoyant Flows},
  author = {Verma, Mahendra K.},
  date = {2018},
  publisher = {{World Scientific}},
  location = {{New Jersey}},
  isbn = {978-981-323-779-7},
  pagetotal = {327}
}

@book{Thorpe2007,
  title = {An {Introduction} to {Ocean} {Turbulence}},
  author = {Thorpe, S. A.},
  date = {2007},
  publisher = {Cambridge University Press},
  location = {Cambridge, UK}
}

@Article{Howard1961,
  author = 	 {L.N. Howard},
  title = 	 {Note on a paper of {John W. Miles}},
  journal = 	 {J. Fluid Mech.},
  year = 	 {1961},
  volume = 	 {10},
  pages = 	 {509--512},
}

@article{Arratia2013,
  title={Transient perturbation growth in time-dependent mixing layers},
  author={Arratia, C. and Caulfield, C.P. and Chomaz, J.-M},
  journal={J. Fluid Mech.},
  volume={717},
  pages={90--133},
  year={2013},
  publisher={Cambridge University Press}
}

@Article{Caulfield2021,
 author      = {C.P. Caulfield},
 title       = {Layering, instabilities and mixing in turbulent stratified flow},
 journal     = {Annu. Rev. Fluid Mech.},
 year        = {2021},
 volume      = {53},
 pages       = {113--145}
 }

@article{newsom2003,
  title={Shear-flow instability in the stable nocturnal boundary layer as observed by {Doppler} lidar during {CASES-99}},
  author={Newsom, Rob K and Banta, Robert M},
  journal={J. Atmos. Sci.},
  volume={60},
  number={1},
  pages={16--33},
  year={2003}
}

@article{bovcek2024kelvin,
  title={{Kelvin-Helmholtz} instability as one of the key features for fast and efficient emulsification by hydrodynamic cavitation},
  author={Bo{\v{c}}ek, {\v{Z}}an and Petkov{\v{s}}ek, Martin and Clark, Samuel J and Fezzaa, Kamel and Dular, Matev{\v{z}}},
  journal={Ultrason. Sonochem.},
  volume={108},
  pages={106970},
  year={2024},
  publisher={Elsevier}
}

@article{liu2015kelvin,
  title={{Kelvin-Helmholtz} instability in an ultrathin air film causes drop splashing on smooth surfaces},
  author={Liu, Yuan and Tan, Peng and Xu, Lei},
  journal={P. Natl. A. Sci.},
  volume={112},
  number={11},
  pages={3280--3284},
  year={2015},
  publisher={National Academy of Sciences}
}

@book{drazin2004hydrodynamic,
  title={Hydrodynamic stability},
  author={Drazin, Philip G and Reid, William Hill},
  year={2004},
  publisher={Cambridge University Press}
}

@article{orszag1971accurate,
  title={Accurate solution of the {Orr-Sommerfeld} stability equation},
  author={Orszag, Steven A},
  journal={J. Fluid Mech.},
  volume={50},
  number={4},
  pages={689--703},
  year={1971},
  publisher={Cambridge University Press}
}

\appendix

\begin{center}
  \large \bf End Matter
\end{center}
\appendix

\section{Appendix A: Derivation of the Self-Similar Stability Equations}
\label{sec:Supp_Stability_Derivation}

\paragraph{Diffusive Non-Dimensionalisation.}
The typical non-dimensionalisation of shear flows, 
such as 
\begin{equation}
\label{eq:supp_oldNonDim} 
\bm{x}_{\textrm{dim}} = \dO \thickspace \bm{x} , 
\qquad 
t_{\textrm{dim}} = \frac{\dO}{\UO} \thickspace t
\end{equation}
used to arrive at equations \eqref{eq:NSE} and \eqref{eq:BE} from the main text \cite{Vieweg2026} and where the subscript $_{\textrm{dim}}$ denotes dimensional variables, relies on the initial finite shear layer half-depth $\dO$ \cite{Arratia2013, Caulfield2021}. However, this approach is obstructive when considering the evolution of an initial step function interface ($\dO \to 0$) as the associated Reynolds number vanishes. 
To resolve this singularity, we propose the alternative scaling 
\begin{equation}
\label{eq:supp_newNonDim} 
\bm{x}_{\textrm{dim}} = \frac{\kappa}{\UO} \thickspace \bm{x}_{\star} , 
\qquad 
t_{\textrm{dim}} = \frac{\kappa}{\UO^2} \thickspace \tstar
\end{equation}
based on the molecular properties (here the thermal diffusivity $\kappa$) of the fluid. This shift ensures the governing parameters remain well-defined as the physical thickness approaches zero. 
The relationship between typical and diffusive coordinates is consequently given by $\left( \bm{x}_\star, \tstar \right) = \ReO \Pr \left( \bm{x}, t \right)$. 

The governing equations translate into
\begin{align}
\nabla \cdot \bm{u}_\star 
&= 0 , \\
\frac{\partial \bm{u}_\star}{\partial \tstar} + \left( \bm{u}_\star \cdot \nabla \right) \bm{u}_\star 
&= - \nabla p_\star + \Pr \nabla^{2} \bm{u}_\star + \Ri_\kappa b_\star \bm{e}_{z}, \\
\frac{\partial b_\star}{\partial \tstar} + \left( \bm{u}_\star \cdot \nabla \right) b_\star 
&= \nabla^{2} b_\star ,
\end{align}
where the diffusion-based Richardson number $\Ri_{\kappa}~:=~\BO \kappa / \UO^3 = \RiO/(\ReO \Pr)$.

\paragraph{Self-Similar Taylor-Goldstein Formulation.}
We decompose the flow into diffusing base states ($U$ and $B$, see also equation \eqref{eq:diffusive_profiles} from the main text) and small perturbations,
\begin{gather}
    \bm{u}_\star(\bm{x}_\star,\tstar) = U(\zstar,\tstar)\bm{e}_{x} + \bm{u}'(\bm{x}_\star,\tstar), \\
    b_\star(\bm{x}_\star,\tstar) = B(\zstar,\tstar)\bm{e}_{x} + b'(\bm{x}_\star,\tstar).
\end{gather}

Applying the self-similar coordinate mapping $\left( \zstar, \tstar \right) \mapsto \left( \xi = \zstar / \sqrt{\tstar}, \tstar \right)$, we substitute the ansatz $\bm{\Phi} \left( \xi, \tstar \right) = \bar{\bm{\Phi}} \left( \xi \right) + \hat{\bm{\Phi}} \left( \xi \right) \tau \left( \tstar \right) e^{ikx}$ (see equation \eqref{eq:ansatz_dif} from the main text) into the linearised equations. This yields the self-similar Taylor-Goldstein generalised eigenvalue problem,
\begin{equation}
    \sigma \left( t \right) \,
    \begin{bmatrix}
        \nabla^2 & 0 \\
        0 & I
    \end{bmatrix}
    \begin{bmatrix}
        \hat{w} \\
        \hat{b}
    \end{bmatrix}
    =
    \begin{bmatrix}
        A_{11} & A_{12} \\
        A_{21} & A_{22}
    \end{bmatrix}
    \begin{bmatrix}
        \hat{w} \\
        \hat{b}
    \end{bmatrix},
\end{equation}
where $D_{\xi} = \partial/\partial\xi$, $\nabla^2 =  D_{\xi}^2/t - k^2$, and the components of the linear operator $\bm{A}$ are defined as
\begin{align}
    A_{11} &= \left( \frac{\xi D_{\xi}}{2 t} + \Pr \nabla^2 \right) \nabla^2 + ik \left( \frac{D_{\xi}^2 U}{t} - U \nabla^2 \right) ,\\
    A_{12} &= -k^2 \Ri_{\kappa} , \\
    A_{21} &= -\frac{D_{\xi} B}{\sqrt{t}} , \\
    A_{22} &= \frac{\xi D_{\xi}}{2 t} + \nabla^2  - ik U .
\end{align}
Note the differences of this system to the traditional approach \cite{Smyth2011}.

To determine the growth rates $\sigma \left( t \right)$ and their corresponding eigenfunctions for a given wavenumber $k$ and time $t$, this generalized eigenvalue problem is discretised in the vertical coordinate $\xi$. The unbounded domain is truncated to a sufficiently large finite extent and the resulting algebraic system is subsequently solved using standard Chebyshev spectral collocation methods \cite{orszag1971accurate} which provide a highly accurate solution to the generalized eigenvalue problem.

\end{document}